\documentclass[runningheads]{llncs}
\usepackage[T1]{fontenc}
\usepackage{graphicx}
\usepackage{color}
\begin{document}
\title{Experimental Implementation of A Quantum Zero-Knowledge Proof for User Authentication}
\author{Marta Irene Garc\'ia-Cid\inst{1,2}%\orcidID{0000-1111-2222-3333} 
\and
Dileepsai Bodanapu \inst{3}%\orcidID{2222-3333-4444-5555}
\and
Alberto Gatto \inst{4}%\orcidID{2222-3333-4444-5555}
\and
Paolo Martelli \inst{4}%\orcidID{2222-3333-4444-5555}
\and
Vicente Mart\'in \inst{2}%\orcidID{0000-0002-2559-3979}
\and
Laura Ortiz \inst{2,5}%\orcidID{2222-3333-4444-5555}
}
\authorrunning{M.I. Garc\'ia-Cid et al.}
\institute{Indra Sistemas S.A., 28108, Madrid, Spain \and
Center for Computational Simulation, Universidad Polit\'ecnica de Madrid, 28660, Madrid, Spain \and
Cohaerentia S.r.l, 20133, Milano, Italy \and Dip. di Elettronica, Informazione e Bioingegneria, Politecnico di Milano, 20133, Milano, Italy
\and
Depto. de Arquitectura y Tecnolog\'ia de Sistemas Inform\'aticos, ETSIInf, Universidad Polit\'ecnica de Madrid, 28660, Madrid, Spain}
\maketitle              % typeset the header of the contribution
\begin{abstract}
A new interactive quantum zero-knowledge protocol for identity authentication implementable in currently available quantum cryptographic devices is proposed and demonstrated. The protocol design involves a verifier and a prover knowing a pre-shared secret, and the acceptance or rejection of the proof is determined by the quantum bit error rate. It has been implemented in modified Quantum Key Distribution devices executing two fundamental cases. In the first case, all players are honest, while in the second case, one of the users is a malicious player. We demonstrate an increase of the quantum bit error rate around $25\%$ in the latter case compared to the case of honesty. The protocol has also been validated for distances from a back-to-back setup to more than $60$ km between verifier and prover. The security and robustness of the protocol has been analysed, demonstrating its completeness, soundness and zero-knowledge properties.  

\end{abstract}
\section{Introduction}
Zero Knowledge Proofs (ZKP) \cite{PrimeraZKP} are cryptographic mechanisms where a user (prover) has to prove to another user (verifier) that the first is aware of a secret, without revealing the secret itself or any information about it. Zero-knowledge proofs provide a powerful tool for enhancing online privacy and security in various domains. Depending on the scope of application, it may be the case that both the verifier and the prover are aware of the secret before carrying out the proof or, on the contrary, that only the prover knows the secret. \\
Depending on the initial setup and the specific needs of the system, a large number of use cases with ZKP applicability can be defined, among which stand out authentication systems to prove the identity of an entity or person \cite{Interactiva2}; privacy-preserving payments to verify that one party has sufficient funds to make a payment, without revealing the actual balance or transaction history \cite{ZKP-Auth}; or access control where users can prove that they have appropriate access to a system or resource without revealing any additional information \cite{ZKP-Access}.\\
The concept of ZKP was first introduced in 1985 by S. Goldwasser, S. Micali, and C. Rackoff \cite{PrimeraZKP} showing that certain types of problems, such as graph isomorphism, could be proven without revealing any additional information beyond the truth of the statement. Since then, ZKP have advanced rapidly, with new techniques, protocols, and applications being developed and refined, becoming an important tool in cryptography. As technologies mature, ZKP are expected to play an increasingly important role in enhancing privacy and security.\\
This type of protocols can be conducted through interactive ZKP \cite{PrimeraZKP}, that is, those in which both the prover and the verifier are required to be present simultaneously during the execution of the proof, as would be the case of the protocol proposed by Fiat-Shamir \cite{Interactiva2}. Non-interactive ZKP can also be implemented \cite{No-interactiva1}, as in the case of the Zero-knowledge succinct non-interactive argument of knowledge (zkSNARK) \cite{zkSNARK}, where the verifier can launch the proof when the prover is absent, thus solving it later. \\
The migration of this type of concepts to the world of quantum communications is of special interest for use cases such as the authentication of several users with access to the same quantum node within a quantum communication infrastructure (QCI). It is important to highlight that this type of user-oriented authentication proposed in this work should not be confused with the authentication of the classic communication channel in the QCI. The latter could solve the authentication of the classic channels used during communications, but there is still a need of guaranteeing the identity of the end user who is on the other side of the screen in such a way that his data remains private. The previous issue is addressed in this work. An example could be the use of the same computer by several doctors in a hospital to upload their patient information into the health system. When accessing the health system each of the doctor must be authenticated. \\
Within the field of quantum cryptography, the Quantum Key Distribution (QKD) provides secret symmetric keys between two remote parties thanks to the fundamental laws of quantum mechanics. The most widely studied and tested QKD protocol is BB84 \cite{BB84}. Recently, other quantum-based cryptographic techniques have been explored, such as quantum digital signatures \cite{QDS1,QDS2} or oblivious transfer \cite{QOT}, among others. \\
This work aims to adapt the concepts of classical zero-knowledge proofs to the field of quantum cryptography, and proposes a new quantum zero-knowledge proof (QZKP). Currently in the literature there are not many studies on QZKP, however some proposals and approaches, mainly for increasing the efficiency of QKD devices, have turned out to be of great interest for the design of the QZKP. Specifically, in 2005 the floating bases protocol was published \cite{bases_infinitas,bases_infinitas_SecProof}, proposing an increase of the number of possible bases to be used in QKD protocols in order to achieve a more efficient system, simultaneously  increasing the threshold of the allowed error rate and reducing the information that can be extracted by Eve. To carry out this scheme, it is required that Bob and Alice have a pre-shared secret key on which the selection of the bases will depend. Another estrategy to improve the efficiency of the QKD devices is the one in \cite{biasedQKD}, where the authors propose a decoy-state protocol for QKD characterized by a biased bases selection, where signal states are always encoded in basis $Z$, while decoy signals can be randomly encoded in basis $X$ or $Z$ with a pre-determined probability. More recently, in 2018, modifications of BB84 protocol were proposed through the use of pseudo-random states generated from a pre-shared secret key \cite{pseudorandomQKD}, in order to achieve higher key rates. The main drawback of the proposed scheme is the strict requirement of the employment of a perfect single-photon source. The use of pre-shared keys and the pseudorandom selection of the quantum states are the main concepts applied for the design of the proposed QZKP.\\
In this paper, we propose and implement a new interactive QZKP where both the prover and the verifier possess a shared secret in advance. The proof is based on purely quantum mechanisms and has been implemented and experimentally tested on quantum cryptographic devices. The paper is organized as follows: firstly, the design of the new QZKP in Section~\ref{QZKP Protocol} is presented; then, the security of the protocol is analyzed in Section~\ref{Security Proof}. Finally, the experimental setup and the outcomes are described in Section~\ref{Implementation}.

%%%%%%%%%%%%%%%%%%%%%%%%%%%%%%%%%%%%%%%%%%%%%%%%%%%%%%%%%%%%%%%%%%%%%%%%%%%%%%%%%%%%%%%
\section{Quantum zero-knowledge proof} \label{QZKP Protocol}
In this paper, we propose an interactive quantum zero-knowledge proof (QZKP), where both the verifier (Alice) and the prover (Bob) must pre-share a secret $s$ to correctly validate the proof. In particular, Bob uses a QZKP with Alice to authenticate himself, as detailed in Figure~\ref{Fig QZKP-Diagram}. This is always the case whenever a QKD channel has been established before. The proof is divided into three stages:

\begin{figure}[ht!]
\centering
\includegraphics[width=1.0\textwidth]{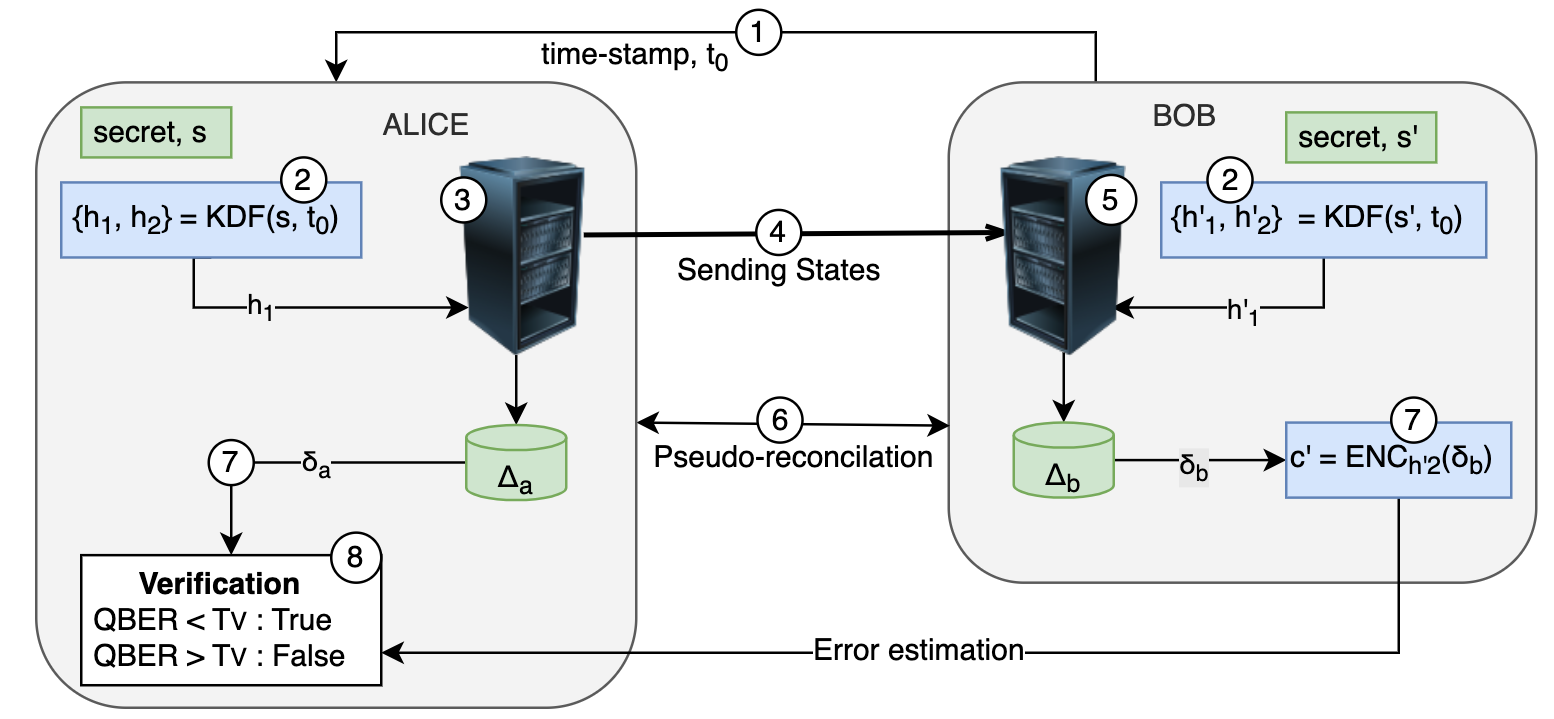}
\caption{Flowchart of the quantum zero-knowledge proof between Alice and Bob. Steps 1 and 2 correspond to the pre-processing stage where the information needed for the execution of the proof is prepared. Steps 3 to 5 correspond to the quantum stage, where the quantum states are prepared, sent and measured. In Steps 6 to 8 the verification of the proof is carried out by the estimation of the quantum bit  error rate (QBER). If both are honest $s=s'$, otherwise $s\neq s'$. KDF means Key Derivation Function; $\Delta_{a,b}$ are raw measurements; $\delta_{a,b}$ are the results of the post-processing of $\Delta_{a,b}$; ENC means the encryption of $\delta_{a,b}$ with $h'_2$; and $T_v$ is the verification threshold.} 
\label{Fig QZKP-Diagram}
\end{figure}

\begin{enumerate}
\renewcommand{\theenumi}{\Alph{enumi}} %Letras mayúsculas
    \item \textbf{Pre-processing stage}, when all the information and setup needed to carry out the QZKP are prepared. The procedures performed in this stage are purely classical and correspond to steps 1 and 2 of Figure~\ref{Fig QZKP-Diagram};
    \item \textbf{Quantum stage}, in which the generation, transmission and measurement of the quantum states will be carried out, permitting to create a raw bit string in both the transmitter and receiver ends. In this case, quantum processes are taking place and correspond to steps 3, 4 and 5 of Figure~\ref{Fig QZKP-Diagram};
    \item \textbf{Verification stage}, in which the validity or not of the proof is determined through an estimation of the error rate. To perform this evaluation, classical tools are used and correspond to steps 6, 7 and 8 of Figure~\ref{Fig QZKP-Diagram}.
\end{enumerate}
  
The pre-processing stage starts with a handshake between Alice and Bob, at the end of which an identical timestamp $t_0$ is generated in both sites. After that, given a secret $s$ of any length, Alice and Bob apply a Key Derivation Function (KDF) \cite{KDF-800-56C} that permits to derive $h_1$ and $h_2$, just giving $t_0$ and $s$ as inputs. This step is carried out by both Alice and Bob simultaneously. The $h_1$ and $h_2$ resulting from this operation on Alice are $h_1 \in \{0,1\}^m$ and $h_2 \in \{0,1\}^n$, where the lengths $m$ and $n$, being $n<m$, are values to be defined by the players before executing the protocol. The process is equivalently performed at Bob's side, where $h'_1 \in \{0,1\}^m$ and $h'_2 \in \{0,1\}^n$ are computed. In a scenario where both Alice and Bob are honest players $s=s'$, otherwise $s\neq s'$. More details about the KDF are given in Section~\ref{Sec:KDF}.

Once $\{h_1, h_2\}$ and $\{h'_1, h'_2\}$ have been calculated, the second phase of the protocol proceeds to generate the quantum bit string. Unlike a conventional QKD protocol, where the bases and states are randomly selected, here the bases are determined on Alice side by the bit-values of $h_1$, while a random selection of states within each basis is performed. Then, Alice sends the stream of quantum signals to Bob, who will receive and measure them following the bases determined by the bit-values of $h'_1$. 

When Bob really knows the secret and Alice is an honest verifier, $h_1=h'_1$. Therefore, when preparing and measuring the quantum states, both will obtain an almost identical bit string, $\Delta^r_a$ in Alice's side and $\Delta^r_b$ in Bob's, meaning with the superindex $r$ that these are raw strings without any post-processing. The obtained bit string is almost identical because, despite preparing the states and measuring them on the same bases, losses occur during their transmission, even in the absence of any malicious manipulation. If we consider an ideal setup, characterized by a perfect extinction ration at the receiver, with no errors in the transmission, neither in the devices, and in absence of eavesdroppers intercepting the communications, the bit string detected by Bob would be identical to the generated one by Alice. 

Once the transmission and measurement of the quantum states is concluded, the verification stage begins with a partial sifting process. This process differs from the standard procedure employed in QKD. In the QZKP, Bob does not publish the bases on which he has measured, because otherwise he would reveal the $h_1$ value related to the secret, which violates the zero-knowledge principle. Instead, he simply announces to Alice the time instants in which he detected a single photon; then, both Alice and Bob selects just these bits from their respective strings, obtaining a sifted string, $\Delta^s_a$ for Alice and $\Delta^s_b$ for Bob, without revealing any information. The superindex $s$ means that these are the sifted strings.

Then, an error estimation between $\Delta^s_a$ and $\Delta^s_b$ is performed, to evaluate the quantum bit error rate (QBER). Typically in QKD, this process is done by both Alice and Bob publishing the same fragment of key as plain text and comparing them. Thus, the number of errors obtained in the selected fragment represents an estimate of the error in the rest of the key. After this comparison, the published fragment is discarded. In QZKP, it is not possible to directly publish a clear fragment of $\Delta^s_a$ or $\Delta^s_b$ because it would reveal an information strongly related to the pre-shared secret $s$. Instead, the selected fragment of $\Delta^s_b$ in Bob, named $\delta_b$ with length $n$, is encrypted with $h'_2$ by means of an One-Time Pad (OTP) procedure, whereby a bit-by-bit XOR is made between $\delta_b$ and $h'_2$, obtaining $c'=ENC_{h'_2}(\delta_b)$. Thus, Bob sends $c'$ to Alice who is able to decrypt it as $ENC_{h_2}(c')=ENC_{h_2}(ENC_{h'_2}(\delta_b))=\delta_b$, if $h_2=h'_2$. Finally, Alice computes the QBER estimation between $\delta_a$ and $\delta_b$. In QZKP, only a rough estimation of the error rate is needed. Errors are neither corrected, nor is privacy amplification performed, since a secret symmetric key is not required at the end of the process. 

Finally, once the QBER has been estimated, the validity of the proof is verified: if the QBER exceeds a certain predefined verification threshold, $T_v$, the proof will give a negative result; on the other hand, if $QBER < T_v$, the proof will be positive, proving the identity of Bob. The QZKP must be performed iteratively $N$ times to guarantee a correct statistical estimate of the QBER.

%%%%%%%%%%%%%%%%%%%%%%%%%%%%%%%%%%%%%%%%%%%%%%%%%%%%%%%%%%%%%%%%%%%%%%%%%%%%%%%%%%%%%%%
\section{Security proof} \label{Security Proof}
\subsection{Security assumptions}
A set of security assumptions must be taken into account during the execution of the protocol.\\ 
First of all, in the different security analysis of the QKD BB84 protocol \cite{BB84_Security,Renner}, a set of assumptions on the adversary are considered that apply equally to the QZKP. Specifically, it is considered that:
\begin{enumerate}
    \item any adversary (external or participant) has unlimited computational power, even with access to a quantum computer;
    \item the quantum channel is considered untrusted;
    \item an external adversary is able to eavesdrop the communication on the classic channel but not to inject messages or modify the content of the information since the channel is assumed to be authenticated.
\end{enumerate}

Moreover, a security perimeter for both Alice and Bob nodes must be guaranteed, in order to avoid any unauthorized physical access to the hardware; as well, appropriate cybersecurity measures are needed to ensure that no side-channels attacks can be performed in both the classical and quantum channels.  

\subsection{Key-Derivation Function details}\label{Sec:KDF}

Regarding the KDF, they are basic and essential components of current cryptographic systems. Their goal is to take some source of initial keying material and derive from it one or more cryptographically strong secret keys. Two types of KDF are defined, according to the standard NIST SP800-56C (r2) \cite{KDF-800-56C}:
\begin{itemize}
    \item One-Step Key Derivation: from a series of inputs, and a secret value, the cryptographic material is derived.
    \item Two-Steps Key Derivation: prior to the derivation, a transformation of the secret is applied.
\end{itemize}
In the QZKP proposed here, a Two-Steps Key Derivation function with a counter mode is recommended \cite{KDF-RFC-5869}. The general structure has two main phases: 
\begin{enumerate}
    \item Extract phase: the keying material ($s$) and a salt value ($t_0$) are taken as input and a fixed-length pseudorandom key $K_{IN}$ is extracted.
    \item Expand phase: the pseudorandom key $K_{IN}$ is expanded into several additional pseudorandom keys ($h_1, h_2$).
\end{enumerate}
Additional input values in the second phase are a label and context, which are fixed values, and the required output length ($m+n$).
It is worth noting that, even though both sources of entropy, $(h_1, h_2)$ are directly derived from the secret $s$, the actual encryption $\delta_a \oplus h_2$  is between two independent elements, since $\delta_a$, even though derived from $s$, is actually built as a random string of bits and therefore independent of $h_2$.

\subsection{QZKP security analysis}
A ZKP has to guarantee completeness, soundness and zero-knowledge. In the following, the security of the proposed QZKP is demonstrated. 

\textbf{Completeness} can be described as: given that both the verifier and the prover are honest and both know the secret, the prover is able to convince the verifier that he does indeed know the secret without revealing it. In the absence of malicious actors, since both parties are aware of the secret, they will get the same bases configuration for states preparation (Alice) or their measurement (Bob). Therefore, in an ideal scenario without photon losses and electronic noise, the results of $\Delta_a^r$ and $\Delta_b^r$ will be perfectly equivalent. Therefore, the estimate of the error would be $QBER=0$ and Bob can fully convince Alice of the knowledge of the secret. However, as we will see in Section \ref{Implementation}, in a realistic implementation, transmission losses and additional sources of noise could be present, increasing the measured the error rate to $QBER\neq0$. 

Assuming that the verifier is honest but the prover is malicious and unaware of the secret, it must be ensured that a dishonest prover is not able to convince the verifier that he knows the secret, except with a negligible probability, to prove the \textbf{soundness} of the proof. These malicious attempts are reflected in the measured QBER, resulting from the execution of the protocol. As proposed by H.-K. Lo analysis \cite{QZKP-SecProof}, the QBER is given by:

\begin{equation}
    QBER = \frac{p_{A,Z}^{2}\cdot e_{B}^Z + p_{A,X}^{2}\cdot e_{B}^X}{p_{A,Z}^{2}+p_{A,X}^{2}} = \frac{(1-r)^2p_{B}^X+r^2p_{B}^Z}{2\left[(1-r)^2+r^2\right]}
\label{eq_QBER}
\end{equation}

Where, $0 < r \leq 1/2$ is a variable parameter which depends on the value of the bits in $h_1$; $p_{A,Z}=(1-r)p_{\mu}$ and $p_{A,X}=rp_{\mu}$ are Alice's probabilities of preparing the states in each basis; $e_{B}^Z=p_{B}^X/2=(1-p_{B}^Z)/2$ the error rate for the case when Alice prepares the state on $Z$ basis and Bob measures on $X$ basis; and $e_{B}^X=p_{B}^Z/2$ the error rate for the case when Alice prepares the state on $X$ basis and Bob measures on $Z$ basis. To try to cheat on Alice, Bob can carry out the following strategy. Given $p_B^Z=0.5$ and $r=1/2$, that is, $50\%$ of the signal states are encoded in the $Z$ basis and $50\%$ in the $X$ basis, since Bob does not know the value of $h_1$, he will measure the signals randomly. In this way he will guess correctly $50\%$ of the times in the selected basis and of the other $50\%$ he will get an uncorrelated result but he will guess correctly the value of the resulting bit half of the times. In total, he will get $75 \% $ of the measurements correct, but without knowing which elements are wrong and which are correct. This strategy raises the QBER to $25\%$ without taking physical errors into account. Therefore, for a $T_v<25\%$ the proof would give a negative result, proving the soundness of the QZKP. This analysis agrees with what is obtained in Eq. (\ref{eq_QBER}) by introducing the parameters.

Finally, if the prover is honest and the verifier a malicious player, the later learns nothing from the proof, demonstrating \textbf{zero-knowledge}. For this aim, Alice could try a similar strategy as in the previous scenario, preparing the quantum states using random bases, since she does not know the secret, $s$. Alice can also try to extract information from the string fragment $\delta_b$ during the error estimation process. However, she does not know the value of $h'_2$, and without this value, it is not possible to correctly decrypt the OTP. If she tries to guess the value of each bit of $c'$, the probability of success guessing all the elements will be: $P_{guess}=1/2^{n}$. In both cases, the obtained QBER will behave similarly as before.\\

%%%%%%%%%%%%%%%%%%%%%%%%%%%%%%%%%%%%%%%%%%%%%%%%%%%%%%%%%%%%%%%%%%%%%%%%%%%%%%%%%%%%%%%
\section{Protocol Implementation} \label{Implementation}

\subsection{Experimental system}
\begin{figure}[ht!]
\centering
\includegraphics[width=1.0\textwidth]{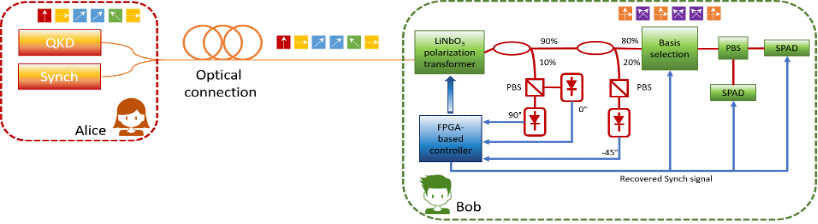}
\caption{Schematics of the pair of discrete-variable quantum cryptographic devices.} 
\label{Fig Experimental_Setup-Diagram}
\end{figure}

The protocol described in Section \ref{QZKP Protocol} has been implemented experimentally exploiting a pair of discrete-variable (DV) quantum cryptographic devices, already tested for standard QKD transmission also in a deployed network in coexistence with classical channels \cite{Turin}. A schematic of the transmitter and receiver is shown in Figure~\ref{Fig Experimental_Setup-Diagram}. 

The DV-QKD prototypes are based on the implementation of the standard BB84 protocol with polarization encoding and decoy-state method \cite{decoy-SecProof}. Alice and Bob exploit a fully-automatic synchronized architecture, thanks to using two different distributed-feedback (DFB) lasers with identical nominal wavelength as sources for the quantum channel and of the auxiliary channel \cite{Dispositivos-Polimi}.  The quantum signal is composed by weak optical pulses with $20~ns$ time duration and $1~kHz$ repetition rate. The selected wavelength is $1310~nm$, useful to avoid Spontaneous Raman scattering photons generated by co-propagating classical sources in the C-band. The decoy-state is implemented as signal ($\mu$), weak decoy-state ($\nu$) and vacuum states ($0$), each of them characterized by a pre-determined probability of occurrence, $p_{\mu}$, $p_{\nu}$, $p_0$, respectively. The measured losses of the receiver module are about $5~dB$. The proposed scheme has been implemented firstly in a back-to-back (B2B) scenario, considering two cases: 1) all actors are honest and 2) the prover is a malicious user who does not know the secret and randomly measures the quantum states received from the verifier.

After the validation of the QZKP in these first, short-distance experiments, the distance between Alice and Bob was increased in order to evaluate the impact on the QBER in the honest condition, ensuring that a relevant number of false positives or false negatives do not occur. For this aim, the QZKP has been tested in a point-to-point standard single-mode fiber (SSMF) link and its performance has been measured, in order to estimate the impact of losses on the QZKP solution. 
The different distances were emulated by inserting optical attenuation in a controlled manner. All the intermediate elements were previously characterized to determine the initial losses introduced by the setup, these being a total of $2.5dB$. Taking into account that the losses in a standard optical fiber are $0.21dB/km$, the setup establishes an emulated initial distance between the devices of $11.9km$. Thus, the evaluated propagation distances ranges from $11.9$~km to $60.6$~km, covering a link attenuation from $2.5$~dB to around $13$~dB.

\subsection{Parameter settings}
In all the experiments carried out for the honest case, the protocol parameters have remained constant except for the length of sifted string $\Delta^s_{a,b}$, named $L_{\Delta}$, that were modified to cover string lengths between $256~bit$ and $2048~bit$. In the case of the length $n$ of $\delta_{a,b}$, the use of $15\%$ of the total of $\Delta^s_{a,b}$ was established for the QBER estimation. All the values of the parameter settings are collected in Table \ref{Tab_Parameter_Settings}.

\begin{table}[]
    \caption{\label{Tab_Parameter_Settings}Parameter settings established during the QZKP executions of the honest and dishonest cases and results of the emulated distance in km, being B2B the back-to-back configuration; the losses in dB; the length $L_{\Delta}$ of $\Delta^s_{a,b}$; the number of iteration that the QZKP has been executed; the average time the system takes for generating $1$ bit; the average QBER estimation; and the standard deviation of the QBER.}
    \centering
    \begin{tabular}{|c|c|c|c|c|c|c|}\hline
         Distance (km) & Losses ($dB$) & $L_{\Delta}$ (bits) & Iterations & Time (s) & QBER & $\sigma_{QBER}$ \\ \hline
         \multicolumn{7}{|c|}{Honest Case} \\ \hline
         B2B & $0$ & $2048$ & $173$ & 0.033 & $0.029$ & $0.007$ \\ \hline
         $11.90$ & $2.50$ & $1024$ & $189$ & 0.077 & $0.028$ & $0.008$ \\ \hline
         $13.62$ & $2.86$ & $1024$ & $858$ & 0.065 & $0.033$ & $0.009$ \\ \hline
         $16.14$ & $3.39$ & $1024$ & $165$ & 0.084 & $0.024$ & $0.009$ \\ \hline
         $17.48$ & $3.67$ & $1024$ & $171$ & 0.089 & $0.029$ & $0.009$ \\ \hline
         $20.19$ & $4.24$ & $1024$ & $612$ & 0.088 & $0.033$ & $0.010$ \\ \hline
         $22.62$ & $4.75$ & $512$ & $229$ & 0.119 & $0.023$ & $0.011$ \\ \hline
         $27.81$ & $5.84$ & $512$ & $10$ & 0.135 & $0.033$ & $0.014$ \\ \hline
         $32.48$ & $6.82$ & $512$ & $10$ & 0.166 & $0.027$ & $0.009$ \\ \hline
         $37.38$ & $7.85$ & $512$ & $10$ & 0.216 & $0.028$ & $0.011$ \\ \hline
         $44.00$ & $9.24$ & $512$ & $11$ & 0.294 & $0.021$ & $0.010$ \\ \hline
         $49.00$ & $10.29$ & $256$ & $11$ & 0.376 & $0.040$ & $0.022$ \\ \hline
         $51.14$ & $10.74$ & $256$ & $11$ & 0.406 & $0.028$ & $0.016$ \\ \hline
         $56.33$ & $11.83$ & $256$ & $10$ & 0.520 & $0.037$ & $0.013$ \\ \hline
         $60.62$ & $12.73$ & $256$ & $26$ & 0.465 & $0.033$ & $0.018$ \\ \hline
         \multicolumn{7}{|c|}{Dishonest Case} \\ \hline
         B2B & $0$ & $2048$ & $190$ & $0.030$ & $0.266$ & $0.015$ \\ \hline
    \end{tabular}
\end{table} 

\begin{figure}[ht!]
\centering
\includegraphics[width=1.0\textwidth]{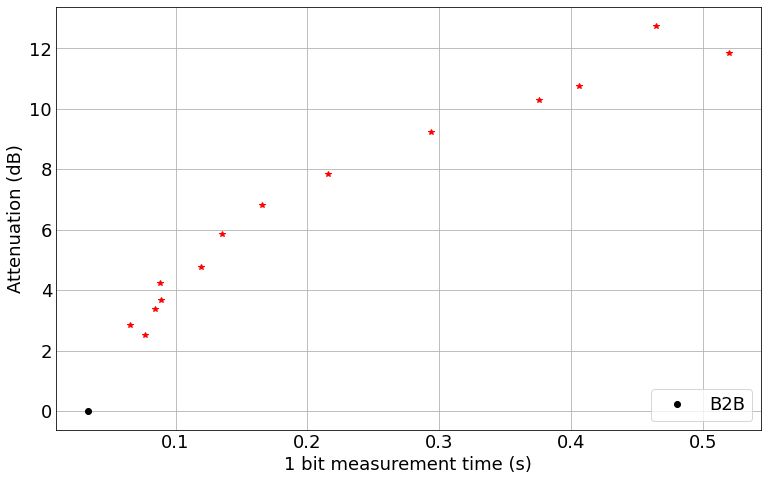}
\caption{Amount of time needed for the generation of $1$ bit in the honest case. The time needed shows a logarithmic behaviour when increasing the losses. The black dot corresponds to the back-to-back (B2B) configuration.} 
\label{Fig 1bit_time}
\end{figure}

The same approach was applied for the execution of the dishonest case but, in this case, the protocol was modified in Bob's side in order to perform random measurements due to the assumption that he ignores the secret, as explained in Section \ref{Security Proof}.
For each experiment, we provide in Table \ref{Tab_Parameter_Settings} the emulated distance in km, being B2B the back-to-back configuration; the losses in dB; the length $L_{\Delta}$ of $\Delta^s_{a,b}$; the number of iteration that the QZKP has been executed; the average time the system takes for generating $1$ bit; the average QBER estimation; and the standard deviation of the QBER. As we can see in Figure \ref{Fig 1bit_time}, the time needed for the generation of $1~bit$ shows a logarithmic behaviour when increasing the losses in the honest case.

\subsection{Analysis of results}
\subsubsection{Comparison between honest and dishonest cases.}
\begin{figure}[ht!]
\centering
\includegraphics[width=1.0\textwidth]{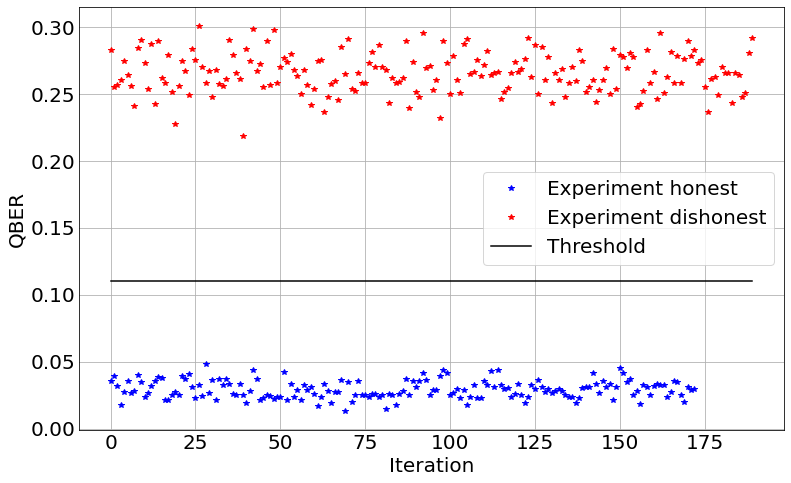}
\caption{Experimental results of the QBER in a back-to-back setup. Blue stars: all players are honest, Red stars: dishonest prover. The black line refers to the standard security threshold value of $11\%$  for the BB84 protocol \cite{BB84_Security}.} 
\label{Fig QBER results}
\end{figure}

For each case, the QZKP procedure has been run for more than $170$ iterations, as shown in Table \ref{Tab_Parameter_Settings}. The outcomes obtained for the average estimated QBER and the standard deviation of the results are shown in Table \ref{Tab_Parameter_Settings} and Figure~\ref{Fig QBER results}.

When verifier and prover are honest (blue stars), the QBER is far below the standard security threshold value of $11\%$ \cite{BB84_Security}. In particular, the measured average QBER shows an error floor of $2.9\%$, owing to the non-idealities of the system, as the finite polarization extinction ratio (ER) of the polarization beam splitter (PBS), limited to $20~dB$, and the presence of dark counts in the two employed single-photon avalanche detectors (SPADs). On the other hand, in presence of a dishonest prover (red stars), the QBER increases up to $26.6\%$, overcoming the security limit of $11\%$ of the BB84 protocol. The QBER performance is stable in time over all the $170$ iterations; some fluctuations are visible both for honest users and dishonest prover, owing to the limited number of bits used for the estimation of the QBER, which is a $15\%$ of $m$; in case of a raw key length of $m=2048$~bits, the estimation is performed on $n=307$~bits. The standard deviations for the two considered configurations are $0.7\%$ and $1.5\%$, respectively. The stronger fluctuation in the dishonest case comes from the limited number of detections and from the statistical behavior of the prover, who measures randomly with equal probability for each basis, while the bases sent by Alice are $N_{\mu}^Z=p_{\mu}\cdot m\cdot (1-r)$ in Z and $N_{\mu}^X=p_{\mu}\cdot m\cdot r$ in X, being $p_{\mu}$ the signal probability, $m$ the length of $h_1$. Anyway, the presence of the fluctuation does not introduce any false positive or false negative condition, permitting to complete the user authentication in all iterations.

\subsubsection{Results over the distance.}
\begin{figure}[ht!]
\centering
\includegraphics[width=1.0\textwidth]{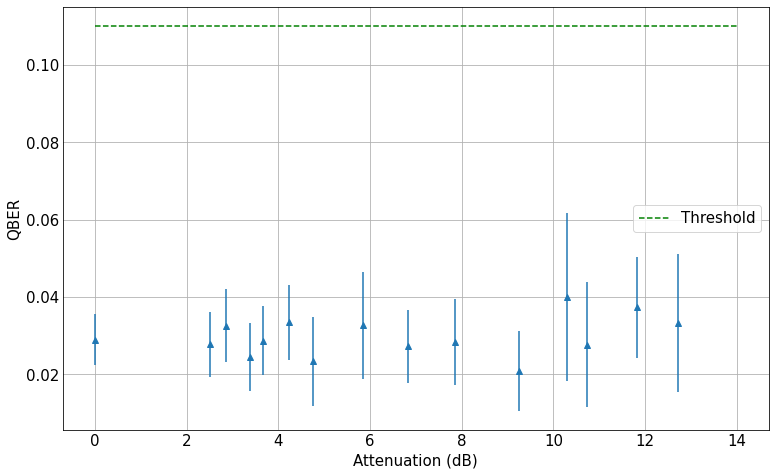}
\caption{Measured QBER performance together with the associated standard deviations versus additional link losses in case of honest parties. The green dashed line refers to the standard security threshold value of $11\%$  for the BB84 protocol \cite{BB84_Security}.} 
\label{Fig QBER distance}
\end{figure}

The measured QBER performance in function of the experimented additional losses is reported in Figure~\ref{Fig QBER distance} and in Table \ref{Tab_Parameter_Settings}.

As before, several acquisitions have been measured for each propagation length, corresponding to several executions of the QZKP. In Figure~\ref{Fig QBER distance}, the average values of the QBER together with their standard deviations are shown. The minimum attenuation at $0~dB$ corresponds to the already described B2B scenario. As can be seen, as the link losses increase, the QBER increases slightly, although it is always far below from the security threshold of $11\%$. As already explained, a minimum error rate close to $3\%$ is present, owing to the limited polarization ER generated by the intrinsic properties of the optical devices and by unavoidable misalignments arising before the PBS. An improvement in the optical components and of the polarization alignment  would reduce the QBER to values less than $1\%$ in the B2B scenario, which would allow us to appreciate with greater detail the gradual increase in QBER with respect to attenuation. In addition, it is observed that the greater the losses, the greater the dispersion of the measured QBER. This is expected due to the limited number of bits used during the QBER estimation influenced by the reduction of the length of $\Delta_a$ when increasing the attenuation of the system, since the measurement time duration required to obtain the established $\Delta_a$ also increases. In the B2B scenario, the generation of $1~bit$ takes an average time of $0.033~s$, with a QBER deviation of $0.7\%$ corresponding to the full protocol execution, while for $12.7~dB$ the generation of $1~bit$ takes an average time of $0.465~s$, giving a standard deviation of $1.8\%$. \\
It is worth to point out that the dishonest case has only be executed in a B2B setup to demonstrate the impact in the QBER when a malicious prover who does not know the secret is present during the execution of the QZKP. This corresponds to the best case scenario for the attacker as there is not additional transmission losses due to the increase in the distance during the proof.

\subsubsection{Comparison between real and estimated QBER.}
\begin{figure}[ht!]
\centering
\includegraphics[width=1.0\textwidth]{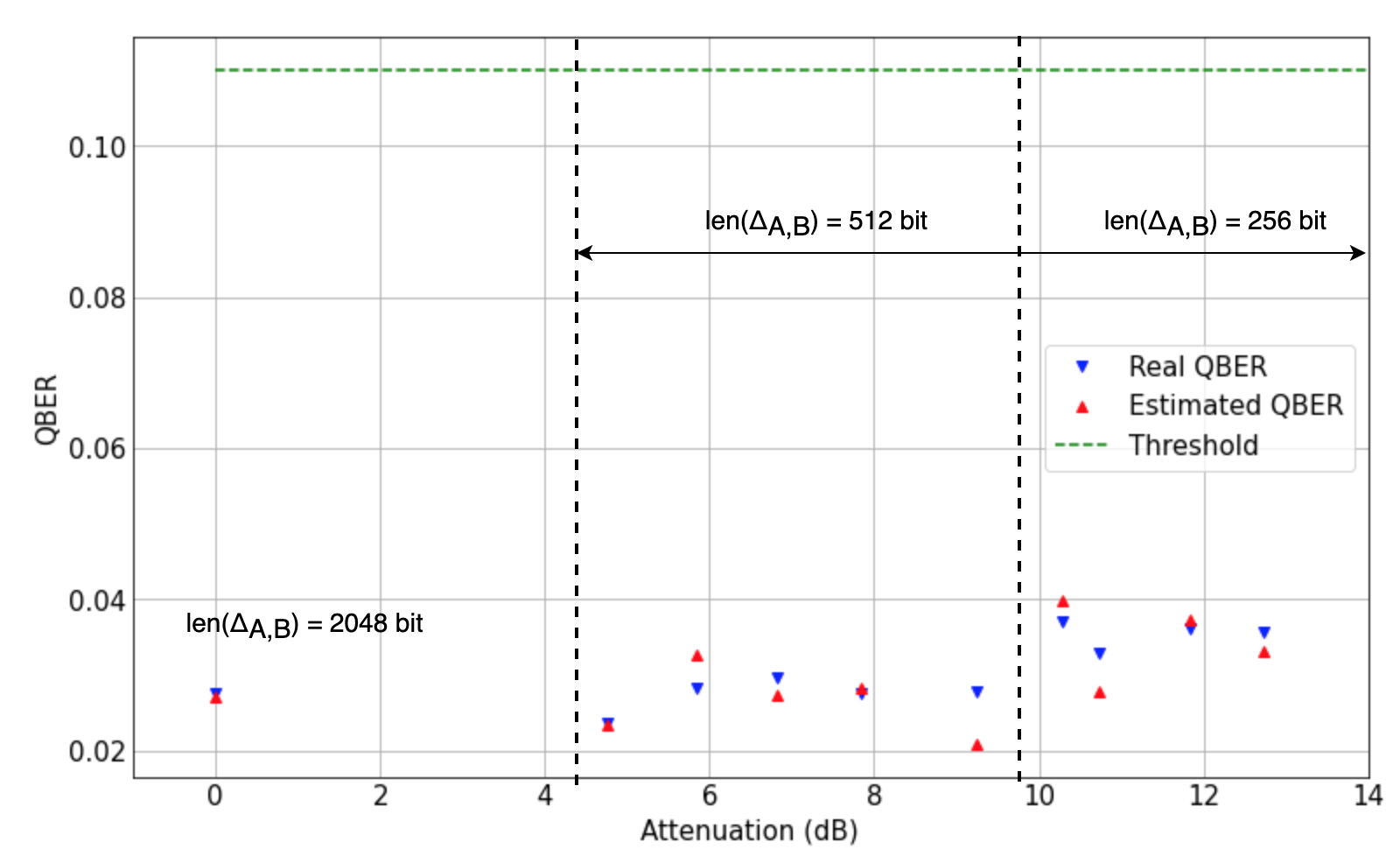}
\caption{Comparison of the real QBER of $(\Delta_a,\Delta_b)$, blue down triangles, versus the estimated QBER obtained from the fragments $(\delta_a,\delta_b)$, red up triangles, for different string lengths.} 
\label{Fig QBER real-estim}
\end{figure}

Finally, given that the QBER used to accept or reject the authentication of a user is an estimate extracted from a fragment of length $n$ of $\Delta_a$ and $\Delta_b$, the variation that exists between this estimate and actual value of QBER has been evaluated for the B2B setup and for raw string outcomes from the largest distances: $22.6~km$ to $60.6~km$. As reported before, the length of the fragment used to estimate the experimented QBER is the $15\%$ of the total segment. To obtain the real value of the QBER, each of the elements of $\Delta_a$ and $\Delta_b$ have been compared bit by bit, obtaining the results showed in Figure~\ref{Fig QBER real-estim} - blue down triangles. 
For its part, the estimation is carried out over the values of $L_{\Delta}$ gathered in Table \ref{Tab_Parameter_Settings}, obtaining the results showed in Figure~\ref{Fig QBER real-estim} - red up triangles.

As we can see, the difference between the estimated value and the real one is less than $1\%$ in the best case at $4.75$~dB and an underestimation of $25\%$ in the worst case at $9.24$~dB, without any negative impact in the authentication test, a behavior that remains constant for all the lengths of $\Delta_{A,B}$ used.

%%%%%%%%%%%%%%%%%%%%%%%%%%%%%%%%%%%%%%%%%%%%%%%%%%%%%%%%%%%%%%%%%%%%%%%%%%%%%%%%%%%%%%%
\section{Conclusions}
To benefit from the advantages provided by QKD, the design of an end-to-end secure cryptographic system is required, where the demonstration of the identity of the two communicating users is one important step to achieve this goal. For this aim, the proposed QZKP is a tool that allows the authentication of users in networks that already have a quantum communications infrastructure without disclosing any personal information about the user or his secret. Based on purely quantum processes the tool provides a quantum-safe authentication mechanism that, not only adds another layer of security to the entire ecosystem, but is also easily implementable with the technology available for QKD and more efficient because it does not require full error correction steps. 
Regarding the security of the protocol, the increase in the order of $25\%$ produced in QBER has been demonstrated, both theoretically and experimentally in a back-to-back scenario, between an honest case, where $QBER=(2.9 \pm 0.7)\%$, and an attempt by a malicious prover to guess the bases associated with the derived $h_1$ function from the secret, where $QBER=(26.6 \pm 1.5)\%$. In addition, the proof is also valid for long distances, being demonstrated for metropolitan areas ($\approx 60~km$), where the increase in the QBER is appreciated as well as a greater dispersion of the data. It is worth to point out that the QZKP has not presented a false positive or false negative, thus demonstrating the robustness of the proof. The QZKP has been tested and guarantees completeness, soundness and zero-knowledge, against different strategies from a malicious player. Finally, we have demonstrated that, for lengths of $2048~bit$, $512~bit$ and $256~bit$, an error estimation using $15\%$ of $\Delta_a$ and $\Delta_b$, provides us with a reliable QBER value that can be used to validate or not the QZKP.

\subsubsection{Acknowledgements} This project has received funding from EIT Digital co-funded by European Institute of Innovation and Technology (EIT), a body of the European Union; the QuantERA II Programme and has received funding from the European Union’s Horizon 2020 research and innovation program under Grant Agreement No 101017733, and State Research Agency – AEI (PCI2022-133009); MCIN with funding from European Union NextGenerationEU (PRTR-C17.I1) and funding from the Comunidad de Madrid - Programa de Acciones Complementarias, Madrid Quantum; and the project “EuroQCI deployment in Spain” (EuroQCI, grant agreement No 101091638). This project has received funding from the European Union's Horizon Europe research and innovation programme under the project "Quantum Security Networks Partnership" (QSNP, grant agreement No 101114043).


\end{document}